# Effects in the Anomalistic Period of Celestial Bodies due to a Logarithmic Correction to the Newtonian Gravitational Potential

Omiros Ragos[1], Ioannis Haranas[2] and Ioannis Gkigkitzis[3]

[1] Dept. of Mathematics, University of Patras, GR-26504 Patras, Greece
e-mail: ragos@math.upatras.gr

[2] Dept. of Physics and Astronomy, York University, 4700 Keele Street, Toronto, Ontario,
M3J 1P3, Canada
e-mail: yiannis.haranas@gmail.com

[3] Departments of Mathematics and Biomedical Physics, East Carolina University
124 Austin Building, East Fifth Street Greenville
NC 27858-4353, USA
e-mail: gkigkitzisi@ecu.edu

**Abstract**
We study the motion of a secondary celestial body under the influence of the logarithmic corrected gravitational force of a primary one. This kind of correction was introduced by Fabris et al. (2009). We derive two equations to compute the rate of change of the periastron w.r.t. the eccentric anomaly and its total variation over one revolution, In a kinematical sense, this influence produces an apsidal motion. We perform numerical estimations for Mercury and for the companion star of the pulsar PSR 1913+16. We also consider the case of the artificial Earth satellite GRACE-A, but the results present a low degree of reliability from a practical standpoint.

**1 Introduction**

In order to explain the difference between the theoretically predicted and the observed position of Mercury's perihelion and its rate of precession, several theories have been proposed. These theories are related to modified versions of the Newtonian potential. Following this direction, Mücket and Treder (1977) introduced a logarithmic correction to the gravitational potential per unit mass. Various authors considered the same potential. Mioc et al. (1991) adopted it in order to estimate the difference between the nodal and Keplerian periods, as well as the changes of the orbital elements over a nodal period. Next, Diacu (1992) examined the validity of the Mücket–Treder gravitational law in the case of a three-body problem. Mioc (2004) has worked out the symmetries of the Mücket–Treder's two-body problem.

In more recent works logarithmic potentials have been used by various researches in investigating the motion of galaxies, the existence and influence of dark matter and the applicability of long-range modified gravity models on the motion of the planets of our solar system. Van Moorsel (1987) found that the data obtained from the observation of the motion of some binary galaxies indicate the presence of dark matter. The influence of this matter can be approximated by a logarithmic potential. Kinney et al. (2001) studied the consequences of adopting the existence of a non-gravitational logarithmic potential instead of that of dark matter in order to explain the discrepancies between the dynamical mass measures of objects such as galaxies and clusters and the observed distribution of luminous matter. Kirillov (2006), while studying the bias relation between visible and dark matter in the case that the structure of the universe does not match that of the Friedman space, he justified that, when a galaxy is near a dark matter point source, a logarithmic-like term should be added to the Newton's potential. Iorio et al. (2008a) worked on the secular precessions of the longitudes of the perihelia of some planets of our solar system and examined if they are compatible with those predicted by long-range modified gravity models. Among others, they studied the results of adopting a logarithmic-type correction to the gravitational potential instead of considering the effect of dark matter. Fabris et al. (2009) analyzed the rotation curves of some spiral galaxies moving within a logarithmically corrected Newtonian potential.

On the other hand, many contributions have been published either on studying the perturbations that affect the orbital elements of celestial bodies or on explaining and modeling the discrepancies between the predictions of the Newton's and/or Einstein's gravitation theory and the available observations on these elements. See, for example, Iorio (2005 ; 2007a), Adkins et al. (2007), Schmidt (2008), Ruggiero (2010), Xu (2011) and Haranas et al. (2011a). Post-Newtonian effects on the anomalistic period have been investigated, too. Iorio (2007b) considered a two-body system in eccentric orbits and examined the post-Newtonian relativistic gravitoelectric part of the precession of the mean anomaly which is not produced by the variation of the orbital period. Li (2010) studied the results of applying three relativity gravitation theories in expressing the post-Newtonian effects in the variation of the periastron passage time for binary stars. Later (2011), he examined the influence of the gravitational radiation damping on this time. Haranas et al (2011b) worked on the effects of a Yukawa-type potential in the anomalistic period of celestial bodies. Last, we must mention that general relativity also predicts the well known gravitomagnetic clock effect in the anomalistic period of a particle orbiting a (slow) spinning main body. This relative literature is certainly too vast to be cited. So, we quote just a review paper written by Iorio et al. (2011a).

In the present work we opt to use a logarithmic correction to the gravitational Newtonian potential in order to calculate the anomalistic time of a secondary body orbiting a primary one. This correction can be expressed as a modification of the Newtonian potential energy per unit mass by the term:

$$V_{\ln}(r) = -\alpha G \mathcal{M} \ln\left(\frac{r}{r_0}\right) \qquad (1)$$

(Fabris et al. 2009; Iorio et al. 2008a) where $\mathcal{M}$ is the mass of the primary, $G$ is the Newtonian gravitational constant, $r$ is the radial distance of the secondary body from the primary one, $\alpha$ is a parameter with dimension of inverse length ($L^{-1}$) and $r_0$ is an arbitrary parameter with dimension of length ($L$). It has been found that a "concordance" value for this parameter is $\alpha \cong -0.1 \, \text{Kpc}^{-1}$. The total acceleration acting on the secondary is:

$$\vec{a}(r) = \left(-\frac{GM}{r^2} + \frac{\alpha GM}{r}\right)\vec{r} = -G(1-\alpha r)\frac{M}{r^2}\vec{r} = -\frac{G_r M}{r^2}\vec{r} \qquad (2)$$

where $G_r = G(1-\alpha r)$. Therefore, the presence of the non-Newtonian term can be considered as converting $G$ into a space-varying Newtonian gravitational constant (Iorio 2011b; Haranas et al. 2011a). In this paper we evaluate our findings using the planet Mercury, the companion star of the pulsar PSR 1913+16, and the artificial satellite GRACE-A. Finally, we compare our results to those obtained by applying a Yukawa-type correction in Haranas et al. (2011b).

## 2 Rate of change and variation per revolution of the periastron time

We consider the unperturbed relative orbit of the secondary body, a Keplerian ellipse. Let $a$ be the semimajor axis, $e$ the eccentricity, $n$ the mean motion, and $M$ the mean anomaly of this orbit. First, we will express the rate of change of the periastron time $T_0$ in terms of the true anomaly $f$. The mean anomaly is defined by

$$M = n(t - T_0) \qquad (3)$$

where $t$ is the time variable. We differentiate Eq. (3) with respect to $t$ and obtain:

$$\frac{dT_0}{dt} = \frac{(t-T_0)}{n}\frac{dn}{dt} - \frac{1}{n}\frac{dM}{dt} + 1. \qquad (4)$$

Using also that, on the unperturbed Keplerian orbit of the secondary, Kepler's third law is given by $GM = n^2 a^3$, the time rate of change of the mean motion is found to be:

$$\frac{dn}{dt} = -\frac{3n}{2a}\frac{da}{dt} \qquad (5)$$

In the presence of a perturbation, the rates of change of the orbital elements can be expressed by means of Gauss' planetary equations: For the semimajor axis and the mean anomaly they read:

$$\frac{da}{dt} = \frac{2}{n\sqrt{1-e^2}}\left[e\sin f\, R + \frac{a(1-e^2)}{r}T\right], \qquad (6)$$

$$\frac{dM}{dt} = n - \frac{(1-e^2)}{nae}\left[\left(\frac{2er}{a(1-e^2)} - \cos f\right)R + \sin f\left(1 + \frac{r}{a(1-e^2)}\right)T\right]. \qquad (7)$$

where $R$ and $T$ are the radial and transverse components of the perturbing acceleration. In our case, $T_{\text{ln}} = 0$ while

$$R_{\text{ln}} = \frac{\alpha GM}{r} = \frac{\alpha n^2 a^3}{r} \qquad (8)$$

Substituting Eqs. (5)–(7) into (4) we obtain that, for $0 < e < 1$:

$$\frac{dT_0}{dt} = \left[\frac{2r}{n^2 a^2} - \frac{(1-e^2)\cos f}{n^2 ae} - \frac{3e(t-T_0)\sin f}{na\sqrt{1-e^2}}\right] R_{\ln} \tag{9}$$

where

$$r = \frac{a(1-e^2)}{1+e\cos f}. \tag{10}$$

Then, we use the well known relations (see, e.g., Murray and Dermott, 1999)

$$r = a(1-e\cos E), \tag{11}$$

$$\frac{dE}{dt} = \frac{n}{1-e\cos E}, \tag{12}$$

$$t - T_0 = \frac{E - e\sin E}{n}, \tag{13}$$

$$\cos f = \frac{\cos E - e}{1 - e\cos E} = \frac{a(\cos E - e)}{r}, \tag{14}$$

$$\sin f = \frac{\sqrt{1-e^2}\sin E}{1-e\cos E} = \frac{a\sqrt{1-e^2}\sin E}{r}. \tag{15}$$

to express Eqs. (8)-(9) in terms of the eccentric anomaly $E$ We obtain that:

$$R_{\ln} = -\frac{\alpha GM}{a(1-e\cos E)} = -\frac{\alpha n^2 a^2}{1-e\cos E} \tag{16}$$

and

$$\frac{dT_0}{dE} = -\alpha \frac{a}{n}\left[2(1-e\cos E) - \frac{(1-e^2)(\cos E - e)}{e(1-e\cos E)} - 3e\left(\frac{E\sin E}{1-e\cos E} + e\frac{\sin^2 E}{1-e\cos E}\right)\right] \tag{17}$$

The logarithmic correction effect on $T_0$ over a whole revolution of the secondary is obtained by integrating $dT_0/dE$ over the interval $[0, 2\pi]$. Then, the change of the anomalistic period per revolution:

$$\Delta T_0 = -2\pi\alpha\frac{a}{n}\left(2 - \frac{1+2e^2}{e^2}\left(-1+\sqrt{1-e^2}\right) - 3\ln\frac{2(1-e)}{1+\sqrt{1-e^2}}\right). \tag{18}$$

If we can measure the change in the anomalistic period per revolution for a given body in an elliptical orbit, we can then write that the coupling constant $\alpha$ is given by the following expression:

$$\alpha = -\frac{ne}{2\pi a}\left(2 - \frac{1+2e^2}{e^2}\left(-1+\sqrt{1-e^2}\right) - 3\ln\frac{2(1-e)}{1+\sqrt{1-e^2}}\right)^{-1} \Delta T_0. \tag{19}$$

## 3 Numerical results

First, we proceed with the calculation of the variation of the anomalistic period of SIRIUS companion αCMaB: Mass of primary = 2.02 $M_{sun}$, semimajor axis of companion $a_B$ = 19.80 AU, (Skemer and Close, 2011) $e = 0.5923$, $n = 3.0 \times 10^{-9}$ rad/s, we obtain that:

$$\Delta T_{\alpha CMaB} = 99.7698 \, \text{s/rev}$$

$\Delta T_{Yuk}$ for Sirius companion

$$\Delta T_{Yuk} = 0.00248014, \, 0.271634, \, 0.28935 \, \text{s/rev}$$

We use lamda and alpha $\lambda = 4.94 \times 10^{15}$ m, and $\alpha = 4.2 \times 10^{-12}, 4.6 \times 10^{-10}, 4.9 \times 10^{-10}$. First, we proceed with the calculation of the variation of the anomalistic period of Mercury. For this planet, we have used the following orbital parameters: $a = 57909083$ km, $e = 0.205$, $n = 8.07 \times 10^{-7}$ rad/s. For the primary, $\mathcal{M} = \mathcal{M}_{\odot} = 1.99 \times 10^{30}$ kg. Applying $\alpha \cong -0.1 \, \text{Kpc}^{-1}$ to Eq. (18), we obtain that:

$$\Delta T_{0 \, Mer} = 4.683 \times 10^{-3} \, \text{s/rev} \tag{20}$$

Next, we estimate the change of the anomalistic period of the companion star of the pulsar PSR 1913+16. The orbital parameters for this star are $a = 1.9501 \times 10^{6}$ km, $e = 0.617$, and $n = 1.575 \times 10^{-4}$ rad/s. The primary's mass is $\mathcal{M} = 1.387 \mathcal{M}_{\odot}$. Then:

$$\Delta T_{0 \, PSRc} = 1.394 \times 10^{-6} \, \text{s/rev} \tag{21}$$

Finally, we calculate the variation of the perigee passage time of the artificial Earth satellite GRACE-A. We have used that, for this satellite, $a = 6876.4816$ km, $e = 0.00040989$ and $n = 0.001100118$ rad/s (http://www.csr.utexas.edu/grace/) and, for the primary, $\mathcal{M} = \mathcal{M}_{\oplus} = 5.9722 \times 10^{24}$ kg. Using Eq. (18) we find that:

$$\Delta T_{0 \, GRACE-A} = 3.184 \times 10^{-10} \, \text{s/rev} \tag{22}$$

We should note here that the degree of reliability of the results concerning GRACE-A is low from a practical standpoint, because of the very small eccentricity. It is known that, for quasicircular orbits, the position of the periastron (hence the periastron time) cannot be accurately determined. However, our results are still of some interest as regards the order of the perigee time variation. A sensitivity analysis for post-Newtonian effects on the GRACE-A and B spacecrafts was worked out by Iorio (2012).

In Figures 1, 2 and 3, we present the variation of rate of change of the anomalistic time w.r.t to the eccentric anomaly $E$ of Mercury, the companion star of the pulsar PSR 1913+16 and GRACE-A. Figure 1 for Mercury indicates that there exist two values of the eccentric anomaly for which this rate is zero. Solving numerically the equation:

$$2(1-e\cos E) - \frac{(1-e^2)(\cos E - e)}{e(1-e\cos E)} - 3e\left(\frac{E\sin E}{1-e\cos E} + e\frac{\sin^2 E}{1-e\cos E}\right) = 0 \tag{23}$$

that results from the R.H.S. of Eq. (17), we obtain that these values are $E = 63.293899°$ and $E = 330.688702°$. The maximum of $dT_0/dE$ can be found by solving:

$$\frac{6e^2 E(e-\cos(E)) - \left(-2 + 6e^2 + 5e^4 - 10e^3\cos(E) + e^4\cos(2E)\right)\sin E}{2e(1-e\cos E)^2} = 0 \tag{24}$$

that comes from the derivative of the L.H.S. of Eq. (23). This maximum is at $E = 215.422386°$. Similarly, for the pulsar PSR 1913+16, $dT_0/dE$ is zero at $E = 32.773697°$ and $E = 359.551488°$ while its maximum is achieved at $E = 306.101830°$. Finally, for GRACE-A, $dT_0/dE$ is zero at $E = 89.929590°$ and $E = 270.070591°$, while its maximum occurs at $E = 180.000091°$. Figures 4 and 5 present the variation of the anomalistic periods of Mercury and the companion star of the pulsar PSR 1913+16 as functions of the eccentric anomaly $E$ and the radial orbital distance $r$.

Several works concerning the effects of Yukawa-type potentials on orbital elements have been published (see, for example, Kokubun 2004; Iorio 2008b). In Haranas et al. (2011), the authors dealt with the anomalistic time change due to such a correction to the Newtonian potential. Comparing the results of the present work with those of the aforementioned publication, we see the following: For Mercury, the change of the anomalistic period because of the logarithmic correction and that by the Yukawa correction with a coupling constant $\alpha_{Yuk} = 3.57 \times 10^{-10}$ are connected through the relation $\Delta T_{0\,\ln} \cong 2.220 \Delta T_{0\,Yuk}$. In the case of the companion star of the pulsar PSR 1913+16 and for $\alpha_{Yuk} = 6.409 \times 10^{-11}$, the corresponding relation is $\Delta T_{0\,\ln} \cong 0.544 \Delta T_{0\,Yuk}$ For GRACE-A and a Yukawa coupling constant in the range $4.2 \times 10^{-12} \leq \alpha_{Yuk} \leq 3.184 \times 10^{-10}$, we obtain that $0.000155 \Delta T_{0\,Yuk} \leq \Delta T_{0\,\ln} \leq 0.0180 \Delta T_{0\,Yuk}$.

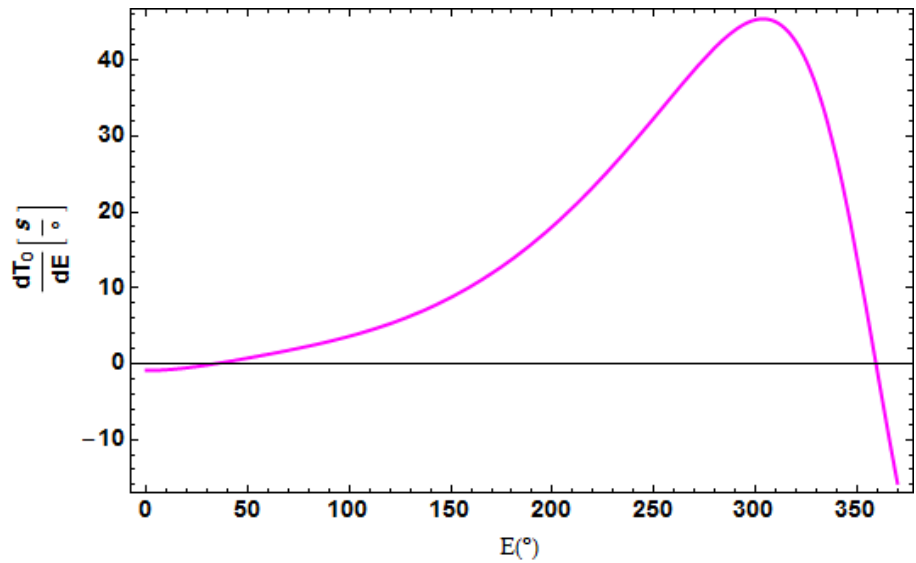

**Fig. 1** Companion of Sirius αCMaB: The variation of the rate of change of the anomalistic period $dT_0/dE$ versus the eccentric anomaly $E$ along a full rotation.

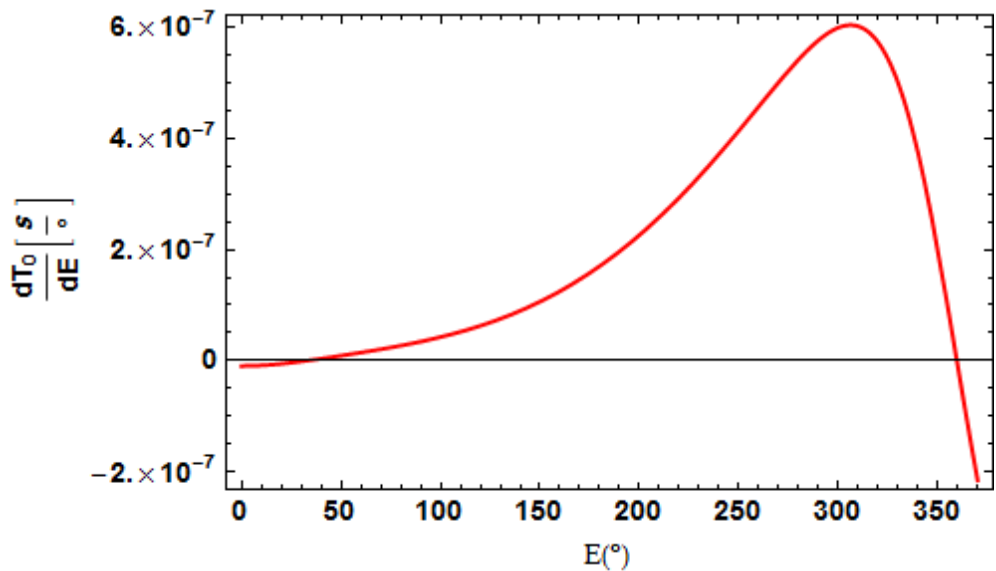

**Fig. 2** Companion star of PSR 1913+16: The variation of the rate of change of the anomalistic period $dT_0/dE$ versus the eccentric anomaly $E$ along a full rotation.

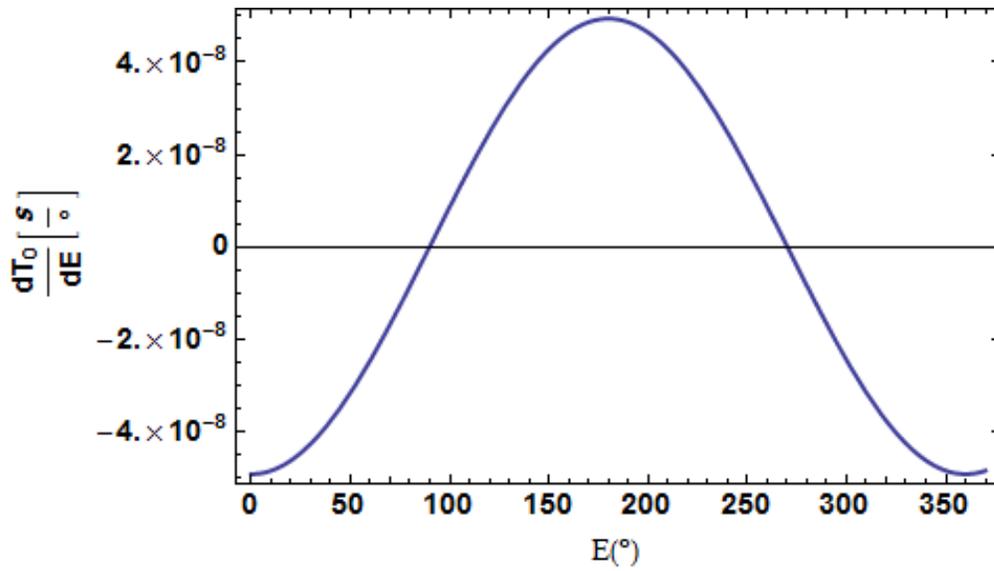

**Fig. 3** Earth's satellite GRACE-A: The variation of the rate of change of the anomalistic period $dT_0/dE$ versus the eccentric anomaly $E$ along a full rotation.

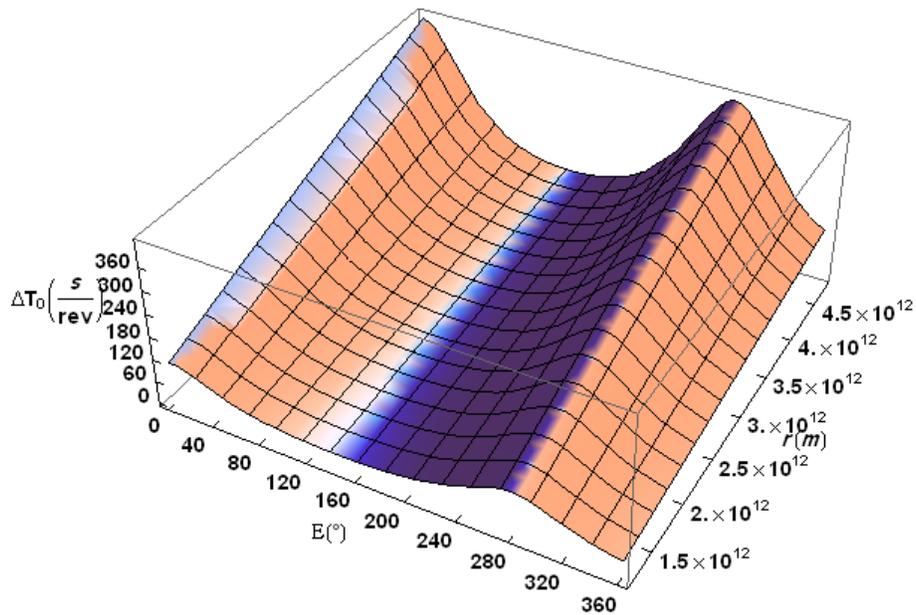

**Fig. 4** Companion of Sirius αCMaB: The variation of the anomalistic period change $\Delta T_0$ versus the eccentric anomaly $E$ and the radial orbital distance $r$ along a full revolution.

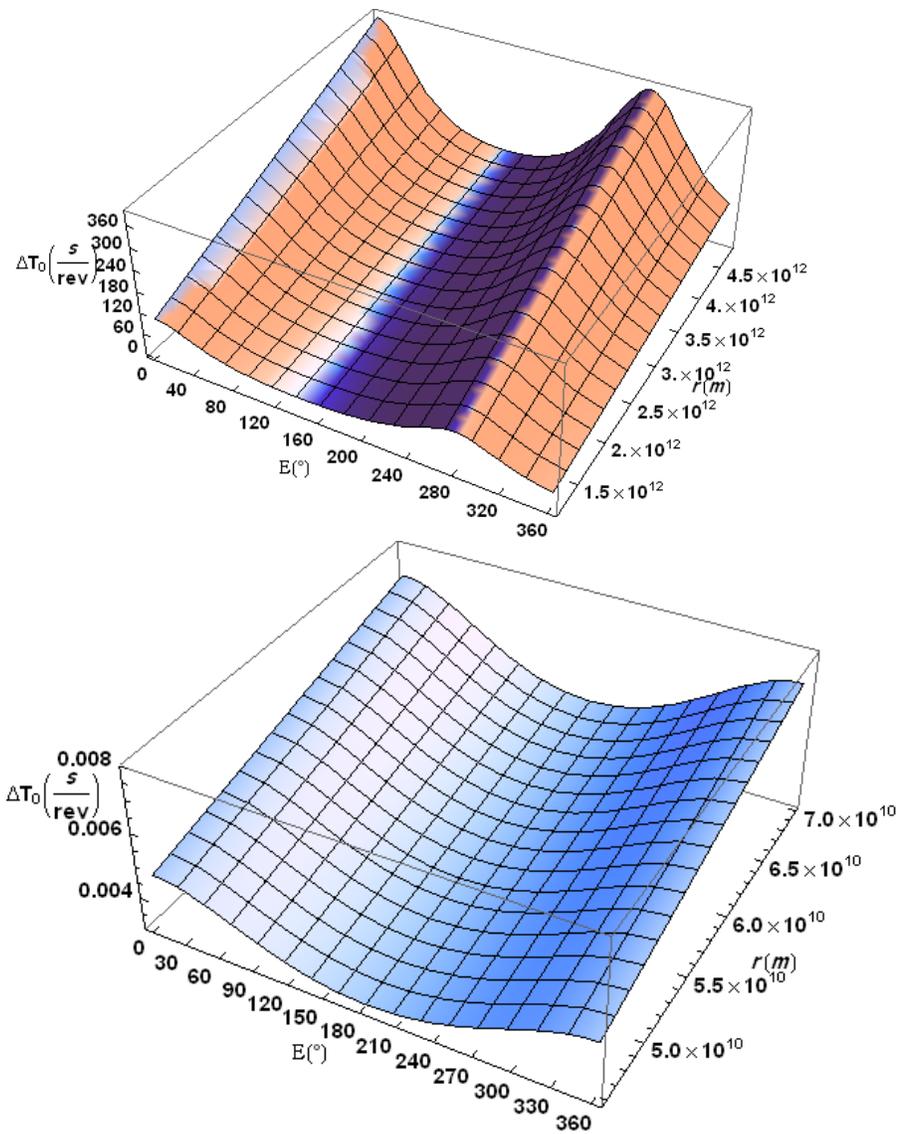

**Fig. 4** Planet Mercury: The variation of the anomalistic period change $\Delta T_0$ versus the eccentric anomaly $E$ and the radial orbital distance $r$ along a full revolution.

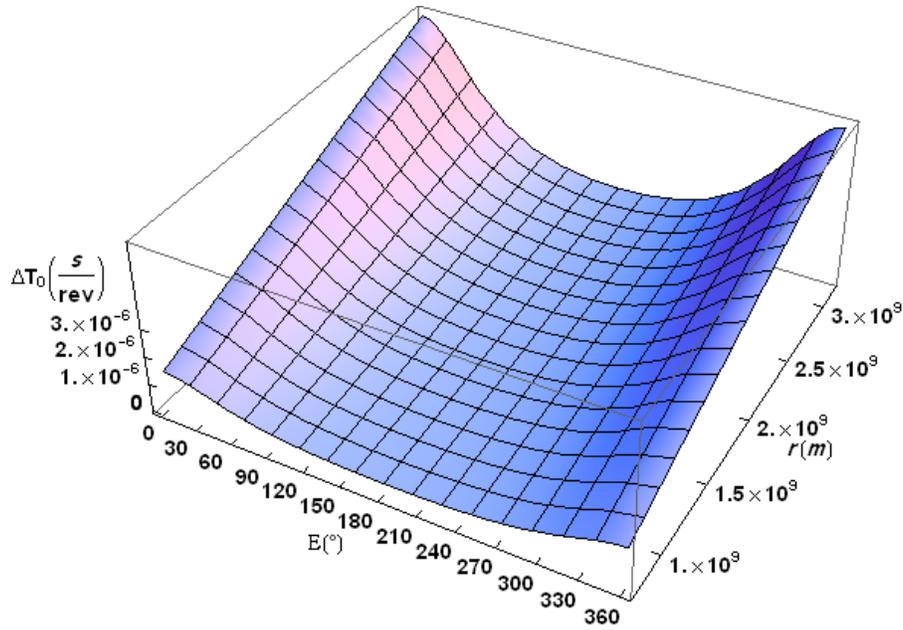

**Fig. 5** Companion star of PSR 1913+16:: The variation of the anomalistic period change $\Delta T_0$ versus the eccentric anomaly $E$ and the radial orbital distance $r$ along a full revolution.

## 4 Summary and concluding remarks

Using a logarithmic correction to the Newtonian gravitational potential as in Fabris et al. (2009) and Iorio et al. (2008a), we derive an eccentric anomaly-dependent equation that estimates rate of change of the anomalistic period of a secondary body orbiting a primary one. By using the integral of this equation over a whole revolution, the contribution of the logarithmic correction to the change of the periastron time can be calculated. This variation was estimated for some concrete astronomical cases. Its observational detection can constitute a possible test for the action of post-Newtonian type forces on the solar system bodies or on other celestial objects. A logarithmic correction is by no means the only kind of correction to be considered in the modification of the Newtonian gravitational potential. For example, general relativistic corrections as well as quantum corrections can be also examined but that is another topic that we are going to deal with in the nearest future.

**Acknowledgements** The authors would like to thank the anonymous reviewer for his valuable comments and suggestions that helped to improve this manuscript considerably.

# References

Adkins, G.S., McDonnell, J.: Orbital precession due to central-force perturbations, Phys. Rev. D **75**(8), id. 082001 (2007).

Diacu, F.N.: On the validity of Mücket–Treder gravitational law, Report DMS-621-IR, University of Victoria, http://dspace.library.uvic.ca:8080/handle/1828/2773 (1992).

Fabris, J.C, Campos, J.P.: Spiral galaxies rotation curves with a logarithmic corrected Newtonian gravitational potential, Gen. Relat. Gravit. **41**(1), 93 (2009).

Haranas, I., Ragos O.: Yukawa-type effects in satellite dynamics, Astrophys Space Sci. **331,** 115 (2011a).

Haranas, I., Ragos, O., Mioc V.: Yukawa-type potential effects in the anomalistic period of celestial bodies, Astrophys. Space Sci. **332**, 107 (2011b).

Iorio, L.: On the effects of Dvali Gabadadze Porrati braneworld gravity on the orbital motion of a test particle, Class. Quantum Gravity **22**(24), 5271 (2005).

Iorio, L.: Astronomical constraints on some long-range models of modified gravity, Advances in High Energy Physics **2007**, Article ID 90731 (2007a).

Iorio, L.: The post-Newtonian mean anomaly advance as further post-Keplerian parameter in pulsar binary systems, Astrophys. Space Sci. **312**(3-4), 331 (2007b).

Iorio, L., Ruggiero, M.L.: Solar system tests of some models of modified gravity proposed to explain galactic rotation curves without dark matter, Scholarly Research Exchange **2008,** Article ID 968393 (2008a).

Iorio, L.: Putting Yukawa-Like modified gravity (MOG) on the test in the solar system, Scholarly Research Exchange **2008**, Article ID 238385 (2008b).

Iorio, L., Lichtenegger, H.I.M., Ruggiero, M.L. Luca; Corda, C.: Phenomenology of the Lense-Thirring effect in the solar system, Astrophys. and Space Sci. **331**(2), 351 (2011a).

Iorio, L.: Observational constraints on spatial anisotropy of *G* from orbital motions, Class. Quantum Gravity **28 (**22), 225027 (2011b).

Iorio, L.: Dynamical orbital effects of general relativity on the satellite-to-satellite range and range-rate in the GRACE mission: A sensitivity analysis, Advances in Space Research **50**(3), 334 (2012).

Kinney, W.H., Brisudova M.: An attempt to do without dark matter, Annals of the New York Academy of Sciences, **927**, 127 (2001).

Kirillov, A.: The nature of dark matter, Phys. Lett. B **632**(4), 453 (2006).

Kokubun, F.: Restricted problem of three bodies with Newtonian + Yukawa potential, International Journal of Modern Physics D **13**(05), 783 (2004).

Li, L.-S.: Post-Newtonian effect on the variation of time of periastron passage of binary stars in three gravitational theories, Astrophys. Space Sci. **327**(1), 59 (2010).

Li, L.-S.: Influence of the gravitational radiation damping on the time of periastron passage of binary stars, Astrophys. Space Sci. **334**(1), 125 (2011).

Mioc, V., Blaga, P.: Orbital motion with the Mücket-Treder post–Newtonian gravitational law, Rom. Astron. Journal **1**(1&2) 103 (1991)

Mioc, V.: Symmetries of Mücket–Treder's two-body problem, Hvar Observatory Bulletin **28(1)**, 167 (2004).

Mücket J.P., Treder, H.J.: The perihelion advance according to a post-Newtonian gravitational law with logarithmic correction term, Astron. Nachr. **298**, 65 (1977).

Murray, C.D. and Dermott, S.F., Solar System Dynamics, Cambridge University Press (1999).

Pitjeva, E. V.: High-precision ephemerides of planets-EPM and determination of some astronomical constants, Solar System Research, **39**(3), 176 (2005).

Ruggiero, M.L.: Perturbations of Keplerian orbits in stationary spherically symmetric spacetimes, eprint arXiv:1010.2114 (2010).

Schmidt, H.-J.: Perihelion precession for modified Newtonian gravity, Phys. Rev. D **78**(2), id. 023512 (2008).

Skemer A., J., and Close, L., M., Sirius B Imaged in the Infrared: No evidence for a remnant planetary system, *Astrophysical Journal*, 730, 53, 2011.

Van Moorsel, G.A.: Dark matter associated with binary galaxies, Astron. Astrophys. **176**(1), 13 (1987).

Xu, Feng: Perihelion precession from power law central force and magnetic-like force, Phys. Rev. D **83**(8), id. 084008 (2011).